\documentclass[pra,twocolumn,superscriptaddress]{revtex4}
\usepackage[]{graphicx}
\usepackage{amssymb,amsmath,multirow}

\usepackage{amsfonts,amssymb,amsmath}
\usepackage[]{graphics,graphicx,epsfig}
\usepackage{amsthm}

\begin{document}


\title{On compression of non-classically correlated bit strings}


\author{Pawe\l\ Kurzy\'nski}
\email{cqtpkk@nus.edu.sg}
\affiliation{Centre for Quantum Technologies,
National University of Singapore, 3 Science Drive 2, 117543 Singapore,
Singapore}
\affiliation{Faculty of Physics, Adam Mickiewicz University,
Umultowska 85, 61-614 Pozna\'{n}, Poland}

\author{Marcin Markiewicz}
\affiliation{Institute of Theoretical Physics and Astrophysics, University of Gda\'nsk, 80-952 Gda\'nsk, Poland}

\author{Dagomir Kaszlikowski}
\email{phykd@nus.edu.sg}
\affiliation{Centre for Quantum Technologies,
National University of Singapore, 3 Science Drive 2, 117543 Singapore,
Singapore}
\affiliation{Department of Physics,
National University of Singapore, 3 Science Drive 2, 117543 Singapore,
Singapore}


\begin{abstract}
We show that the outcomes of measurements on correlated quantum systems that are spatially separated can be compressed much more efficiently than their classical counterparts. We show this on an example of bit strings generated by singlet correlations that we compress using Huffman coding. We then draw general conclusions on compressibility of quantumly correlated strings using Kolmogorov complexity.

\end{abstract}


\maketitle


{\it Introduction.---} Understanding correlations present in physical systems is crucial. This starts at a level of two particles with the phenomenon of quantum entanglement \cite{Schrodinger}, quantum teleportation \cite{teleportation}, super-dense coding \cite{super-dense} to name a few and ends with with various fundamental phenomena in complex systems consisting of a large (usually Avogadro) number of particles such as phase transitions, quantum phase transitions \cite{phase}, super conductivity and many others. 

Since realisation that information is physical \cite{landauer}, correlations carried by physical objects have been utilised to perform computational tasks. In this context, two qubit correlations have been deployed, amongst many other applications, to perform device independent quantum key distribution \cite{Ekert91} and private randomness amplification \cite{Acin}, and multi-qubit correlations are used to perform quantum computation \cite{cluster}.

Compression of information plays a pivotal role in information processing tasks as shown by Shannon in his groundbreaking paper \cite{Shannon}. The achievements of our digital era are significantly based on the fact that classical information can be efficiently compressed and stored. Quantum information, which could be the future of computing, can also be compressed \cite{Schumacher-comp} and stored \cite{Horodecki}. 

The ultimate limits of compression can be identified with the help of algorithmic entropy also known as Kolmogorov complexity (KC) \cite{CT}. KC of a bit string $a$ of length $n$, $K(a)$, is the length of the shortest programme on a universal Turing machine that generates $a_n$. The string is called complex if its Kolmogorov complexity is large in comparison to $n$ and it is called simple otherwise. KC is essential in our understanding of interplay amongst computing, information and randomness \cite{Vitanyi}.  

An appealing albeit rather philosophical meaning of KC is that of quantification of our knowledge about physical processes. Any physical process can be viewed as generation of bit strings and a physical theory as an attempt to predict these strings. A successful physical theory, i.e., a theory with predictive power, can be viewed as a universal Turing machine with a programme that can efficiently describe all possible strings one can observe in experiments that are thought to be in the domain of its validity. The programme should be much shorter than the strings that it generates (otherwise the theory has no predictive power) and thus it should be simple according to KC.

In this paper we investigate compressibility of information generated by two spatially separated parties, Alice and Bob, who share an entangled state of two qubits. Both perform some number of binary non-commuting measurements on their respective qubits generating finite strings of bits. This is a primitive of quantum correlations because it involves the smallest quantum systems for which correlations are non-trivial. We are interested in how well Alice's and Bob's bit strings can be compressed in comparison to a scenario where Alice and Bob share classically correlated random variables.

This paper is inspired by ideas of Zurek \cite{Zurek}, Schumacher \cite{Schumacher}, Bennett {\it et. al.} \cite{Bennett} and Cilibrasi, and Vitanyi \cite{CV}.


{\it Compression of correlated bit strings.---} The scenario we consider consits of two spatially separated parties Alice and Bob sharing a singlet state of two qubits $|\psi_-\rangle=(|01\rangle-|10\rangle)/\sqrt{2}$. Each of them performs a measurement of a qubit along one of the $N$ fixed directions on the Bloch sphere. Each direction is chosen randomly. We denote Alice's measurement directions by $X_1, \dots, X_{N}$ and Bob's $Y_1, \dots, Y_{N}$. After many measurements they can arrange their data into binary strings each of length $n$ denoted by $x_i, y_j$. In this notation $x_i$ denotes a string that was assembled from Alice's qubit measurement outcomes for the setting $X_i$; $y_j$ is Bob's string assembled from the outcomes for the setting $Y_j$. The spatial separation and random choice of measurement directions is to enforce certain constraints on Alice's and Bob's strings that we will discuss soon.

We now ask a question: how well Alice and Bob can loslessly compress their pairs of strings $x_i$ and $y_j$? Since these strings originate from measurements on a singlet state, they are localy fully random and hence $x_i$ and $y_j$ are un-compressible when considered separately. However, Alice can compress her data provided that she has some information about Bob's data (and vice versa). This compression is possible because of correlations between their bit strings.

In principle, the ultimate compression rate Alice and Bob can achieve is given by the Shannon entropy $S(x_i y_j)$, however this rate can only be achieved in the limit of infinitelly long bit strings or if the probabilities of finite strings are powers of two \cite{CT}. In all other cases one has to use some more practical compression algorithms. Since the work of Shannon many compression algorithms that loslessly compress finite data were invented. In this work we apply the Huffman prefix coding algorithm \cite{CT}.

The compression procedure goes as follows. First, Alice produces the bit string $z_{ij}=x_i \oplus y_j$, i.e., she adds Bob's bit string $y_j$ to her bit string $x_i$ (modulo two). Next, she uses the Huffman code to compress $z_{ij}$ to some shorter string $\tilde{z}_{ij}$. The string $y_j$ is not compressed. This is the losless compression, since the Huffman coding can be reversed $\tilde{z}_{ij} \rightarrow z_{ij}$ and $z_{ij}\oplus y_j=x_i$.

Before we proceed, let us give the reader an idea on how the Huffman coding algorithm is applied in our scenario by considering a specific example. Imagine that Alice generated the following bit string with the help of Bob's data $$z_{ij}=000010010000001100.$$ The length of $z_{ij}$ is $n=18$. Alice divides her bit string into $m$ parts, each consisting of $k=n/m$ bits. In this example we choose $k=2$, however different choices of $k$ lead to different compression efficiencies $$00|00|10|01|00|00|00|11|00.$$

Next, Alice counts the frequency of each 2-bit sequence: $\#00=6$, $\#01=1$, $\#10=1$, $\#11=1$. She makes a table in which 2-bit sequences are ordered from the most frequent to the less frequent $$\{(00,6),(01,1),(10,1),(11,1)\}.$$ The first term in the bracket corresponds to the sequence, whereas the second to its frequency. She follows the recursive algorithm --- she starts at the end of the table and combines the two last elements into a new element whose frequency is the sum of frequencies of these elements. Next, she produces a new table in which the last two elements are superceded by the new one. Note, that the new element can take some higher position in the table because its frequency may not be the smallest. In our case the new table is $$\{(00,6),([10,11],2),(01,1)\},$$ where $[10,11]$ denotes that the new element is composed of $10$ and $11$. Alice repeats the whole procedure to obtain $$\{(00,6),([[10,11],01],3)\},$$ and after one more round she obtains $$\{([[[10,11],01],00],9)\}.$$

\begin{figure}[t]
\begin{center}
\includegraphics[width=4cm]{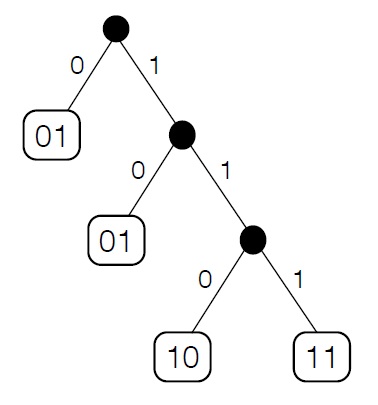}
\end{center}
\vspace{-0.5 cm}
\caption{\label{f1} The Huffman tree for the four two-bit sequences. The code for a specific sequence is constructed by going down from the root to the node corresponding to the element. If we go to the left we add 0 to the code word and if we go to the right we add 1. For example, the code word corresponding to 10 is 110.}
\end{figure}

The element $[[[10,11],01],00]$ is used to produce the Huffman tree from which we obtain the Huffman prefix-free coding (for details see Fig.~\ref{f1}): $00 \rightarrow 0$, $01 \rightarrow 10$, $10 \rightarrow 110$ and $11 \rightarrow 111$. Therefore, after compression we get $$\tilde{z}_{ij}=00110100001110.$$ The length of $\tilde{z}_{ij}$ is $\tilde{n}=14$ and the compression rate is $r_{ij}=\tilde{n}/n=7/9\approx 0.778$.

It is also instructive to estimate the Shannon entropy of the string $z_{ij}$. This can be done under an assumption that each bit in $z_{ij}$ was generated independently and with respect to the same probability distribution. In this case we estimate the  probabilities of $0$ and $1$ from frequencies, i.e., $p(0)=7/9$ and $p(1)=2/9$. These probabilities give the Shannon entropy $S(z_{ij})\approx 0.764$, which shows that our Huffman compression is reasonably efficient.

The next step is to evaluate the expected Huffman compression rates for bit strings $z_{ij}$ that are generated from quantum correlations. Firstly, we note that the subsequent pairs of bits in $x_i$ and $y_j$ are generated by independent pairs of qubits in the singlet state. The probability that Alice's measurement along direction $\vec{a}$ gives an outcome $x$ and that Bob's measurements along $\vec{b}$ gives $y$ ($x,y\in\{0,1\}$) is given by
\begin{equation}
p(x,y|\vec{a},\vec{b})=\frac{1}{4}\left(1-(-1)^{x+y}\vec{a}\cdot\vec{b}\right).
\end{equation}
Therefore, the probabilities of $0$ and $1$ in the bit string $z_{ij}$ are given by
\begin{equation}
p(0|z_{ij})=\frac{1}{2}\left(1-\vec{a}\cdot\vec{b}\right),~~p(1|z_{ij})=\frac{1}{2}\left(1+\vec{a}\cdot\vec{b}\right).
\end{equation}

Following the previous example, after $n$ rounds of measurements Alice produces $z_{ij}$ from her string of outcomes $x_i$ and Bob's string of outcomes $y_j$. She divides this string into $m$ substrings of size $k=n/m$. Each substring belongs to the set of all $2^k$ possible bit strings. Now we ask, what the
expected compression rate is? The expected frequency for the substring containing $\#0=l$ and $\#1=k-l$ is $p(0|z_{ij})^l p(1|z_{ij})^{k-l}m$. We can plug these frequencies into a similar table we used before and construct the Huffman code that will give us an expected compression rate.

Note that the algorithm does not depend on $m$. The only important parameters are $k$ and the probabilities $p(0|z_{ij})$ and $p(1|z_{ij})$. Interestingly, even in the case of extremal probabilities $p(0|z_{ij})=0$ and $p(1|z_{ij})=1$, or vice versa, the compression rate depends on $k$. In this case every k-bit sequence can be coded as one bit and the length of $\tilde{z}_{ij}$ is $n/k$. Thus, the compression rate $r_{ij}=1/k$.

We finish our discussion of the compression algorithm with yet another example. The expected compression rate for measurements for which $\vec{a}\cdot\vec{b}=1/\sqrt{2}$ and for $k=2$ is $r_{ij}\approx 0.709$. This rate is better for larger $k$. For $k=4$ it is $\approx 0.611$ and for $k=8$ it is $\approx 0.605$. The corresponding Shannon entropy is $S(z_{ij})\approx 0.601$. Note, that up to now we have not exploited the non-classicality of quantum correlations. We do this in the next section.


{\it Compression of non-classically correlated bit strings.---} In this section we present our main result. We derive an inequality for compression rates of classically correlated bit strings. Then we show that this inequality is not obeyed by bit strings generated by some non-classical correlations such as those generated by a singlet state of two qubits.

In order to derive this inequality we use the properties of NCD \cite{CV}. NCD of two bit strings $x_i$ and $y_j$ is defined as
\begin{equation}
NCD(x_i,y_j)=\frac{C(x_i,y_j)-\min\{C(x_i),C(y_j)\}}{\max\{C(x_i),C(y_j)\}},
\end{equation}
where $C(x_i), C(y_j)$ are compressed sizes of $x_i$ and $y_j$ respectively. $C(x_i,y_j)$ is the compressed size of the joint (concatenated) bit string $x_i y_j$. The compression $C(x_i)$ can be some real-world compression algorithm (for instance, Huffman or gzip) but it can also be KC $K(x)$ or Shannon entropy $S(x)$ since both functions bound efficiency of realistic compressors \cite{CV,Bennett,Zurek}.

In case of the singlet state local bit strings are fully random and are non-compressible, hence for any compressor $C(x_i)=C(y_j)=n$. On the other hand, we showed that the compression of the concatenated bit string $x_i y_j$ can be done via compression of $z_{ij}=x_i\oplus y_j$ and leaving $y_j$ uncompressed. Therefore, $C(x_i y_j)=C(z_{ij}) + n$ and for the singlet state NCD simplifies to
\begin{equation}
NCD(x_i,y_j)=\frac{C(z_{ij})}{n}=r_{ij}.
\end{equation}

Let us go back to the scenario in which Alice and Bob, each, have $N$ measurements. In order to proceed we need to make a set of assumptions that characterise behaviour of classical strings in the context of the experiment discussed here. (i) We assume that the bit strings $x_i$ and $y_i$ are uniformly complex, i.e., any n-bit subset of those strings is equally compressible. This assumption can be verified experimentally. (ii) Because of the spatial separation of Alice and Bob and the finite speed of light compression rate of the string $x_i$ alone does not depend on which measurement was chosen by Bob. If this was not the case Bob could send superluminal signals to Alice, which is not compatible with the special relativity theory (iii) Compression rate of strings $x_i$ and $y_j$ is the same as of the strings $x_i'$ and  $y_j$. Here $x_i'$ denotes a string generated for $i$th setting of Alice but not at the same time as the string $y_j$.  This is a counterfactual statement that cannot be tested experimentally although it has been extensively used in the literature on EPR paradox and Bell inequalities \cite{review}. We need one more assumption, which is also of counterfactual nature (iv) The triangle inequality for any distance measure must be obeyed even for the strings that cannot be simultaneously generated \cite{Triangles}. For instance, $NCD(y_1,y_N)$ cannot be determined experimentally because the strings $y_1$ and $y_N$ come from measurements of incompatible observables.

The basic building block of our construction is the fact that every distance metric obeys the triangle inequality. NCD obeys the triangle inequality up to a factor that depends on the length of the uncompressed bit string \cite{CV}. We start with the triangle inequality between the bit strings $x_1, y_N,y_1$ (first two strings appear on the left-hand side of the inequality, convention we use throughout the paper)
\begin{equation}\label{t1}
NCD(x_1,y_{N}) \leq NCD(x_1,y_1) + NCD(y_1,y_{N}) + O\left(\frac{\log n}{n}\right).
\end{equation}
Next, consider another triangle inequality between the strings $y_1,y_N,x_2$.
Combining these two triangle inequalities we get a "rectangle" inequality
\begin{eqnarray}
NCD(x_1,y_N) &\leq&  NCD(x_1,y_1) + NCD(x_2,y_1)  \nonumber \\ &+& NCD(x_2,y_N) + O\left(\frac{2\log n}{n}\right).
\end{eqnarray}

We follow analogical steps until we are left with terms $NCD(x_i,y_i)$ or $NCD(x_{i+1},y_i)$. Finally
\begin{eqnarray}
NCD(x_1,y_N) &\leq&  \sum_{i=1}^N NCD(x_i,y_i) + \sum_{i=1}^{N-1} NCD(x_{i+1},y_i)  \nonumber \\ &+& O\left(\frac{N\log n}{n}\right). \label{iq}
\end{eqnarray}
From now on we assume sufficiently long bit strings $N/n \ll 1$ such that the last term can be omitted. Since classically correlated bit strings obey all triangle inequalities used in this derivation, they also have to obey (\ref{iq}).

However, we now show that bit strings generated by the singlet state correlations violate the inequality (\ref{iq}). Let us consider the measurement settings $\vec{a_i}$ and $\vec{b_j}$
\begin{eqnarray}
\vec{a_i}&=&\left( \sin(i-1)\theta, ~0, ~\cos (i-1)\theta \right), \nonumber \\
\vec{b_j}&=&\left( \sin (j-1/2)\theta, ~0, ~\cos (j-1/2)\theta\right),
\end{eqnarray}
where $\theta=\pi/(2N-1)$. This choice of directions yields $\vec{a_i}\cdot\vec{b_i}=\vec{a}_{i+1}\cdot\vec{b_i}=\cos\frac{\pi}{4N-2}$ and $\vec{a_1}\cdot\vec{b_N}=0$, which implies $r_{ii}=r_{i+1,i}=r$ and $r_{1N}=1$. The uncompressability of $x_1 y_N$ follows from the lack of correlations between $x_1$ and $y_N$ because of the orthogonality of the corresponding Bloch vectors.

For the singlet state correlations the inequality (\ref{iq}) simplifies to
\begin{equation}\label{iq2}
\frac{1}{2N-1} \leq r.
\end{equation}
It is violated for $N \geq 3$. In particular, for $N=3$ and $k=9$ we get $r\approx 0.199$. For $k=10$ we get $r\approx 0.192$. For comparison, the Shannon limit in this case is $S(z)\approx 0.166$. The reason why this time our compression rate is not as close to Shannon rate as in the previous examples is that the Huffman coding is not optimal if a probability of some k-bit sequence is close to 1. In such cases one needs to choose higher values of $k$.

We now show that the violation of the inequality (\ref{iq}) implies violation of the inequality
\begin{eqnarray}
Z(x_1,y_N) &\leq&  \sum_{i=1}^N Z(x_i,y_i) + \sum_{i=1}^{N-1} Z(x_{i+1},y_i)  \nonumber \\ &+& O\left(\frac{N\log n}{n}\right). \label{iq2}
\end{eqnarray}
$Z$ is Zurek's distance measure \cite{Zurek} defined for two bit strings of length $n$ as $Z(a,b)=2K(a,b)-K(a)-K(b)$, where $K(a)$ is KC of the string $a$ etc.
This inequality can be derived in exactly the same way as (\ref{iq}). Its violation stems from 1) $K(x_i)=K(y_j)=n$ because individual strings by Alice and Bob are purely random 2) $K(x_iy_j)\leq C(x_iy_j)$, i.e., Kolmogorov complexity is optimal by definition 3) The strings $x_1$ and $y_N$ are not correlated and as such cannot be efficiently compressed, i.e., $K(x_1\oplus y_N)=n$.  To prove it, we observe that the left-hand side of the inequality (\ref{iq2}) equals one whereas the right-hand side is bounded from above by the right-hand side of the inequality (\ref{iq}), which ends the proof.

{\it Discussion.---} We have shown that bit strings generated by entangled qubits violate information-theoretic inequalities that are obeyed by classically correlated random variables. The violation occurs because the assumptions used to derive the inequalities (\ref{iq}) and (\ref{iq2}) are not satisfied by quantum mechanical correlations in the presence of entanglement. A natural question is to identify the unfulfilled assumptions from the set (i) - (iv). It is clear that if one rejects a possibility of infinite speed of information propagation in the universe then it must be the assumption (iii) or (iv) that is not obeyed in quantum theory. We would like to point out that all the assumptions (i)-(iv) are satisfied by classically correlated random variables.

The interesting conclusion based on the violation of the inequalities (\ref{iq}) and (\ref{iq2}) is that quantum mechanics (equivalently, violation of the assumptions (iii) and (iv)) allows for more efficient compression than classical correlations. We can also conclude that Kolmogorov complexity is not a proper description of all possible  correlations occurring in nature, thus giving a possibility to distinguish between quantum and classical correlations on the level much more fundamental that the probabilistic one.

{\it Acknowledgements.---} P. K. and D. K. are supported by the Foundational Questions Institute (FQXi) and by the National Research Foundation and Ministry of Education in Singapore. M. M. is supported by the International PhD Project ``Physics of future quantum-based information technologies'' grant MPD/2009-3/4 from Foundation for Polish Science and  by the University of Gda\'nsk grant 538-5400-B169-13.



\end{document}